\documentclass{jfm}
\usepackage{graphicx}
\usepackage{epstopdf, epsfig}

\usepackage{graphicx}
\usepackage{amsmath}
\usepackage{enumitem}
\usepackage{siunitx}
\usepackage{bm}
\usepackage{floatrow}
\usepackage{color}
\usepackage[labelformat=simple]{subcaption}

\usepackage{url}
\usepackage[colorlinks=true,citecolor=blue,linkcolor=blue]{hyperref}

\newfloatcommand{capbtabbox}{table}[][\FBwidth]

\renewcommand{\vec}[1]{\boldsymbol{\mathrm{#1}}} 
\newcommand{\ten}[1]{\boldsymbol{\mathrm{{#1}}}}
\providecommand\bnabla{\boldsymbol{\nabla}}

\shorttitle{The Brinkman viscosity for porous media exposed to free flow}
\shortauthor{A. Rinehart, U. L\={a}cis and S. Bagheri}

\title{The Brinkman viscosity for porous media exposed to a free flow}

\author{Aidan Rinehart,
  U\v{g}is L\={a}cis
 \and Shervin Bagheri\corresp{\email{shervin@mech.kth.se}}}

\affiliation{FLOW centre, Department of Engineering Mechanics KTH, SE-100 44 Stockholm, Sweden}

\begin{document}

\maketitle

\begin{abstract}
The Brinkman equation has found great popularity in modelling the interfacial flow between free fluid and a porous medium. However, it is still unclear how to determine an appropriate effective Brinkman viscosity without resolving the flow at the pore scale.
Here, we propose to determine the Brinkman viscosity for rough porous media from the interface slip length and the interior permeability. Both  slip and   permeability can be determined from unit-cell analysis, thus enabling an \textit{a priori} estimate of the effective viscosity. By comparing the velocity distribution in the porous material predicted from the Brinkman equation with that obtained from pore-scale resolved  simulations, we show that modelling errors are $\sim 10\%$ and not larger than $40\%$. We highlight the physical origins of the obtained errors and demonstrate that the Brinkman model can be much more accurate for irregular porous structures. 
\end{abstract}

\begin{keywords}
...
\end{keywords}

\section{Introduction} \label{sec:intro}

Flows over porous surfaces are encountered in a wide range of natural and industrial settings. For example, organisms 
use porous layers to gain unique capabilities such as self-cleaning, insect trapping, and enhanced locomotion \citep{Fish_YANG2014,Dolphin_Kramer1962,Dolphin_Carpenter2000,Shark_Bandyopadyay2014,Worm_Zhao2018}.  The features of the interface that separates the overlaying free-flowing fluid and the porous layer play a vital role in many applications such as soft tissue generation, nutrient exchange, chemical reaction and filtering \citep{Tissue_Podichetty2014,Stephanopoulos1989,Grenz2000}. 
Observations of these naturally occurring phenomena and the many interesting emerging applications have recently lead to extensive efforts to model the interaction between porous media and free
flows~\citep{carraro2013pressure,carraro2018effective,lacis2016framework,lacis2020transfer,Angot2017,nakshatrala2019interface,zampogna2019modeling,bottaro2019flow,eggenweiler2021effective}.

The challenge of modelling the flow over porous material stems from describing the exact pore-scale geometry of the solid material and corresponding influence on flow above it. Porous structures in nature span a wide range in complexity and functionality~\citep{liu2011bio}. The range of length scales encountered in the porous structures and associated flow within the material can become prohibitive for direct numerical 
simulations~\citep{keyes2013multiphysics}. Effective models relax this requirement
and provide a macroscopic description of the porous media that in turn depends on the macroscopic properties of the material.
For modelling single phase fluid flow through porous material, the
Brinkman model,
\begin{equation}
\vec{u} = - \frac{\ten{K}}{\mu_f} \cdot \bnabla p + 
\frac{\mu_b}{\mu_f}\,\ten{K} \cdot \Delta \vec{u}, \label{eq:Brinkman}
\end{equation}
is a common choice. Here, $\vec{u}$ and $p$ are macroscale flow and pressure fields, $\ten{K}$ is the permeability
tensor, $\mu_f$ is the fluid viscosity and $\mu_b$ is the Brinkman or effective viscosity. We define the ratio of Brinkman and fluid viscosities as $\mu = \mu_b / \mu_f$. This model was originally proposed by
\citet{Brinkman} as an empirical extension of seminal
Darcy's law~\citep{darcy1856fontaines}.
Darcy's law has been derived using various theoretical approaches \citep{whitaker1998method,mei2010homogenization} as a leading order description of seepage in bulk porous media.
The Brinkman equation
has also been obtained through
statistical formulations~\citep{Tam1969,Lundgren1972}.
In multi-scale expansion, the Brinkman term appears as a second-order
correction~\citep{Auriault2005}. 


The Brinkman term, in theory, should be neglected for porous materials with realistic solid content. However, the Brinkman model has remained a popular choice for practitioners
working on turbulent flows~\citep{rosti2015direct,gomez2019turbulent},
tissue modelling~\citep{Tissue_Podichetty2014},
and porous gels~\citep{feng2020boundary}. The main reasons
behind the wide adoption are the practical 
advantages of the Brinkman model. These are the ability to accommodate 
the no-slip condition at the wall and straightforward coupling with
standard Navier-Stokes equations.
The theoretically sound
Darcy's law has an intrinsic slip at the solid boundaries and
also requires more complex coupling with free
fluid~\citep{lacis2016framework,lacis2020transfer,sudhakar2021msebc}.
However, the practical advantages of the Brinkman model are tied
with two issues. First, despite the straightforward coupling,
there is no universal choice for the physical form of boundary
conditions. The existing options range from stress
continuity~\citep{Marty1994,bottaro2016rigfibre}, different types
of stress jump~\citep{Angot2017,lu2019optStBrinkm} to a combination of velocity
and stress jumps~\citep{lacis2020transfer,sudhakar2021msebc}.
The second issue is an uncertainty of the Brinkman viscosity $\mu_b$, which also is the main focus of the current work.

It is generally accepted that in the limit of small ratio between solid volume and total volume, $\Phi$, the Brinkman viscosity should equal the
fluid viscosity \citep{Auriault2009}.
This has often led
 practitioners of the Brinkman equation to assume
$\mu = 1$~\citep{Stephanopoulos1989,Tissue_Podichetty2014,gomez2019turbulent}.
However, Brinkman himself wrote~\citep{Brinkman} that ``the factor
$\mu_b$ ... may be different than $\mu_f$''.
For larger $\Phi$ one encounters a vast range of
the Brinkman viscosity values.
Levy and Auriault~\citep{Levy1983,Auriault2005,Auriault2009}
argued that
the assumptions behind the Brinkman equation derivation
limit its application to exceedingly sparse porous domains.
As such, the interaction between particles would be negligible and
the Brinkman viscosity should always equal the
fluid viscosity.
\citet{Angot2017} proposed that the Brinkman equation
is valid for all solid volume fractions
and the Brinkman term would simply become negligible for
dense porous domains.
Most theoretical attempts to determine
the effective viscosity~\citep{Brinkman,Koplik1983,Ochoa-Tapia_1995,Starov2001,freed1978}
have focused on sparse porous materials.

\begin{figure}
    \centering
    \includegraphics[trim = 0 140 0 120,clip,width=\textwidth]{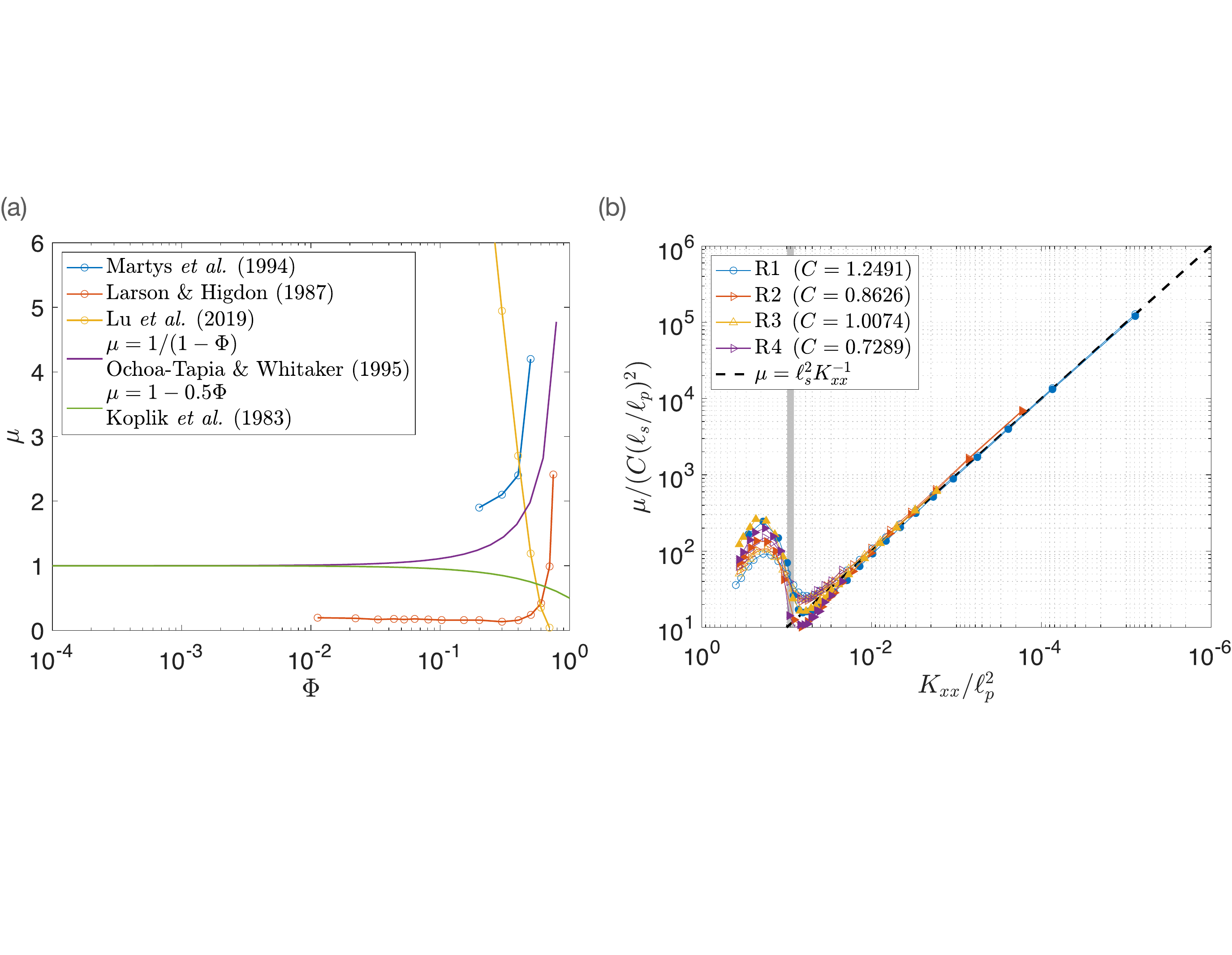}
    \caption{Literature review for effective viscosity as a function of solid volume fraction (a). Empty circles indicate numerical studies and solid lines theoretical derivations. Scaling for effective viscosity $\mu$ obtained within this study (b). R1-R4 refer to different rough porous geometries (figure~\ref{fig:PorousModel}e, table~\ref{tab:geom-summary}). We report $\mu$ for both sheared driven (empty symbols) and pressure-driven (filled symbols) flows.
        In (b), vertical grey rectangle correspond to the region where the solid volume fraction of all
        geometries increases beyond $\Phi = 10^{-2}$.}
    \label{fig:lit_mu}
\end{figure}


Several authors have obtained the Brinkman viscosity by matching  pore-scale simulations with the Brinkman equation for large $\Phi$. \citet{Larson1986,Larson1987} computed the pore-scale flow for regular circular cylinders and determined $\sqrt{\mu}$ from the associated decay rates. They investigated  axial and transverse flows and found that the square root of effective viscosity is bounded as $0.5 < \sqrt{\mu} < 2$ in solid volume fraction range $0 < \Phi < 0.8$.
%
\citet{lu2019optStBrinkm} obtained optimal $\mu$ values by matching the pore-scale simulations with the solution from coupled Stokes-Brinkman problem. They found $\mu$ as a decreasing function of $\Phi$.
\citet{Marty1994} computed the flow in porous media built using randomly overlapping spheres for pressure and shear driven cases. They obtained $\mu$ always greater than unity and observed decaying towards unity for decreasing $\Phi$. It was also stated that ``good fits'' were obtained for $0.2 < \Phi < 0.5$. However, a quantitative characterisation was lacking. \citet{kolodziej2016} simulated Couette flow in domains filled with regular circular cylinders. They found $\mu$ always less than unity. More recently, \citet{zaripov2019} investigated pressure-driven porous channel flows containing either square or circular cylinders. This is the first study we are aware of where a quantitative evaluation of the Brinkman equation accuracy is reported alongside the effective viscosity. They also demonstrate that a parameter $\lambda^2 = k\,\mu$, where $k$ is a measure of permeability, remains roughly constant over a wide range of solid volume fractions. However, \citet{zaripov2019} only consider pressure-driven flow in the interior of porous material and compute the error in velocity flux -- a single integral measure for each simulation.

It is clear 
that the value of Brinkman viscosity, in particular for configurations consisting of porous material and free fluid, is still an openly debated question.
The currently proposed $\mu$ values from the literature
are
presented in figure~\ref{fig:lit_mu}(a) with expressions and authors contained in the legend.
The reported $\mu$ values span a wide range and it is unclear how to choose $\mu$.
%
%
%
In this work, it is proposed to obtain
the effective viscosity $\mu$ through a
scaling estimate.
To estimate $\mu$, we focus on the Brinkman term
alone and
consider a configuration driven only by
shear ($\bnabla p = 0$). From
equation~(\ref{eq:Brinkman}) we observe that the Brinkman term
must balance the Darcy dissipation term. We write
\begin{equation}
u_c \sim \mu\,k\,\frac{u_c}{\ell^2}, \label{eq:intro-est}
\end{equation}
where $u_c$ is some characteristic velocity
and $\ell$ is some characteristic
length.
This estimate has been previously employed by
\citet{Auriault2009} to discuss the validity of the Brinkman
equation. \citet{Auriault2009} concluded that the Brinkman equation is valid for vanishing solid volume
fractions. Contrary, we
show that this estimate can be
useful also for large solid volume fractions.
From the estimate (\ref{eq:intro-est}), it can be argued that the effective viscosity
$\mu$ should be expressed as
\begin{equation}
\mu = C \frac{\ell^2}{k},
\label{eq:scale}
\end{equation}
where $C$ is dimensionless constant of order unity,
as will be discussed later.
%
We will verify this viscosity estimate by matching pore-scale resolved simulations with predictions from the Brinkman model.
In theory, for the estimate (\ref{eq:scale}) to be fully compatible with a macroscale description,
the length scale $\ell$ should be a macroscopic length scale. However,
we set the length scale to $\ell = \ell_s$,
where $\ell_s$ is the Navier-slip
length~\citep{navier1823memoire,lacis2020transfer,sudhakar2021msebc}
of the porous material. This yields a good collapse of
the Brinkman viscosity for all the rough porous geometries we have
considered, see figure~\ref{fig:lit_mu}(b). As a result,
the proposed scaling (\ref{eq:scale}) can be a useful tool for obtaining the effective viscosity for
a given practical problem. However, the slip length $\ell_s$ is itself
microscopic quantity. 
Therefore the Brinkman
model in this parameter regime should be viewed as
an ad-hoc model. Consequently, a practitioner has to take into
account the accuracy limitations, which we also demonstrate
in this work.

The paper is organized as follows. In \S\ref{sec:BrinkVisc}, we describe the flow configuration used both for pore-scale resolved simulations and the effective simulations. In \S\ref{sec:Surface}, we describe porous surface classification in smooth and rough ones.
In \S\ref{sec:Disc}, we provide guidelines for applying the Brinkman equation to practical problems, including discussion about interface conditions.
Finally, we provide concluding remarks in \S\ref{sec:Conc}.

\section{Matching direct numerical simulations with the Brinkman model}\label{sec:BrinkVisc}

To obtain the effective viscosity $\mu$, we consider a one-dimensional channel flow. We investigate the flow
driven by moving top wall, as well as the flow driven by pressure gradient along the channel.


\subsection{Pore-scale resolved flow fields}

\begin{figure}
\includegraphics[trim = 0 80 0 170,clip,width=0.95\textwidth]{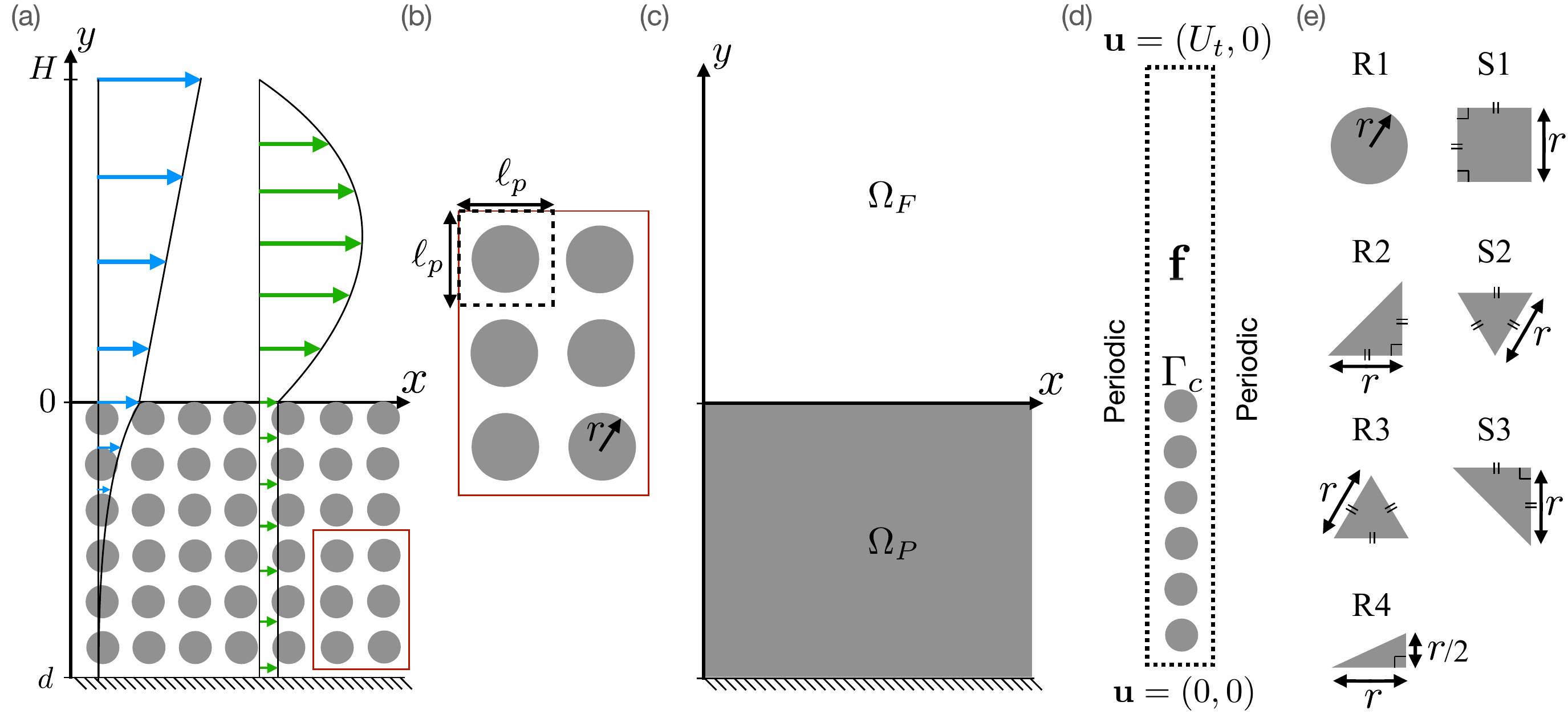}%
\caption{
The flow geometry used in this work (a). Simulation domains in the Brinkman model (c). Close-up of porous structure (b), where dashed lines indicate REV boundaries. Domain for resolved Stokes simulations (d). Geometries considered in this work (e).
}
\label{fig:PorousModel}
\end{figure}

The chosen flow geometry is shown in figure~\ref{fig:PorousModel}(a).
The porous domain is constructed from representative
element volumes (REV) depicted with dashed lines in figure~\ref{fig:PorousModel}(b). The solid volume fraction $\Phi$ for a porous
material is defined as
\begin{equation}
    \Phi = \frac{a}{\ell_p^2},
    \label{eq:SVF}
\end{equation}
where $a$ is the solid cylinder cross section area and $\ell_p^2$ is the area of the REV.
The circular cylinder area is $a = \pi r^2$, where $r$ is the circle radius.
For the other porous geometries the appropriate cross-section area, $a$, is used in equation (\ref{eq:SVF}). We investigate seven different porous geometries as shown in figure~\ref{fig:PorousModel}(e) and summarized in table~\ref{tab:geom-summary}.

We assume that the flow is sufficiently slow such that it can be described by the 
Stokes equations,
\begin{align}
    \nabla p = \mu_f \Delta \vec{u} + \vec{f}, \\
    \nabla \cdot \vec{u} = 0,
\end{align}
where $\vec{f}$ is a volume force.
Solutions to the Stokes equations are computed using the finite element code FreeFem++~\citep{freefem}. The velocity is solved with quadratic (P2) elements and pressure with linear (P1) elements. The computational domain is defined by a single column of the porous structure as shown in figure~\ref{fig:PorousModel}(d). A single column represents a complete solution since the flow structures are smaller than the computational domain.  Periodic boundary condition is imposed on the vertical boundaries of the domain. Velocity $\vec{u}$ is prescribed at the upper boundary. No-slip condition is applied to 
the cylinder surfaces ($\Gamma_c$) and bottom of the domain.
For the shear driven flow, $U_t = U$, $\vec{f} = (0,0)$ while for pressure driven flow, $U_t = 0$, $\vec{f} = (\mu_f U / \ell_p^2,0)$. Here,
$U$ is a characteristic velocity scale.
%
Pore-scale solutions were checked for mesh dependence. Doubling the mesh resolution resulted in effective viscosity values changing by less than 2\%.

\begin{figure}
    \centering
    \hspace*{-40pt}
    \includegraphics[trim = 0 80 0 80,clip,width=1.15\textwidth]{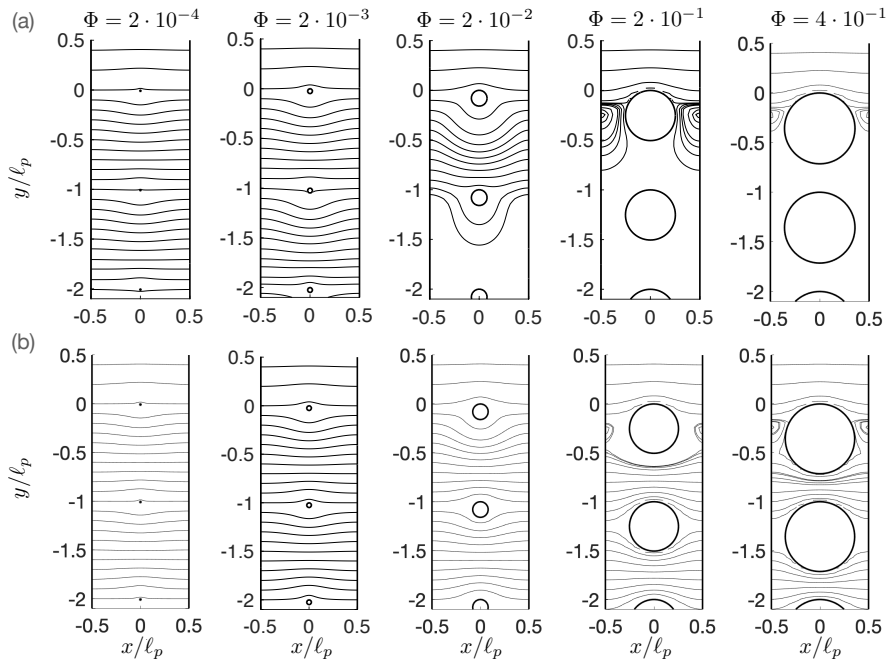}
    \caption{Streamlines around circular cylinders in shear-driven (a)
     and pressure-driven (b) configurations.}
    \label{fig:Streamlines}
\end{figure}

In figure~\ref{fig:Streamlines}, we show streamlines of the pore-scale solutions for several $\Phi$.
Streamlines are shown for velocity larger than $10^{-5}\,U$. For small $\Phi$ only minor deflections of streamlines are observed. Increasing $\Phi$ leads to larger curvature in the streamlines.
The curved streamlines illustrate that both velocity components exist in pore-scale flow.
Recirculation zones are observed for both driving mechanisms.
In the shear driven flow (figure~\ref{fig:Streamlines}a) a rapid velocity decay is observed, in particular for the largest two solid volume fractions. In the pressure-driven flow (figure~\ref{fig:Streamlines}b), on the other hand, 
there is a finite Darcy flow inside the porous material.



\begin{figure}
    \centering
    \includegraphics[trim = 0 100 0 50,clip,width=1.0\textwidth]{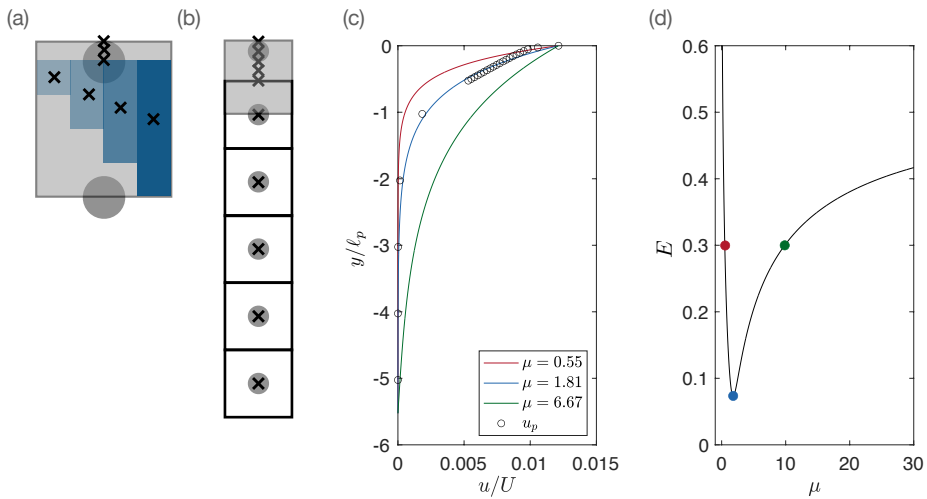}
    \caption{Averaging of DNS and comparison to Brinkman model. Blue rectangles (a) represent volume average windows near the interface. Black crosses (a,b) show vertical locations of averaged data points from the DNS. Velocity profiles (c) for three effective viscosities (solid lines) and averaged DNS $u_p$ (open circles), circular cylinders, $\Phi = 0.002$. (d) The error \eqref{eq:L2error} of Brinkman velocity profile as a function of $\mu$.}
    \label{fig:viscosity_example}
\end{figure}

For a fair comparison between pore-scale (DNS) resolved velocity
and the Brinkman model, the DNS velocity field must be averaged.
The appropriate averaging of the pore-scale flow field is still an open question.
In the current work, we choose to use a combination of plane and volume averages.
In the interior of the domain, we compute volume averages over
each REV. This is analogous to how Darcy's law is derived through
homogenisation~\citep{mei2010homogenization} or
the method of volume-averaging~\citep{whitaker1998method}.
However, the flow velocity undergoes rapid change near the free-fluid interface.
To capture this, we have chosen to gradually transition from
the volume average in the interior to the plane average near the 
free fluid.
The transition is implemented through a sequence of continuously shrinking volume averages.
The location of the last volume averaged data point is exactly in
the midpoint between the first and second solid cylinder. The location of the first plane averaged data point is at the centre
of the uppermost solid cylinder.
In figure~\ref{fig:viscosity_example}(a,b), we show the location
and shape of each averaging window.
%
Obtained averaged DNS data points $u_p$ are shown with open circles in figure~\ref{fig:viscosity_example}(c). 


\subsection{Velocity distributions from Brinkman model}



To obtain the velocity distribution by solving the Brinkman equation \eqref{eq:Brinkman}, we need
values of permeability tensor and effective viscosity.
We compute the permeability tensor
on a single REV by using a unit-cell problem \citep{whitaker1998method} exposing the REV (figure~\ref{fig:PorousModel}b) to volume forcing along all coordinate directions. The effective viscosity $\mu$ is left as a free parameter.

Streamlines of the DNS (figure~\ref{fig:Streamlines}) are symmetric with respect
to the vertical centre line. Therefore the average vertical velocity
is zero.
%
This results in a 1D macroscopic flow configuration.
The Brinkman equation (\ref{eq:Brinkman}) simplifies to a second-order ordinary differential equation
(ODE) for the streamwise velocity,
\begin{equation}
u_x = - \frac{K_{xx}}{\mu_f} p_x + \mu K_{xx} \frac{d^2 u_x}{dy^2}, \label{eq:1d-ode-poreFlow}
\end{equation}
where $p_x$ is an externally imposed constant driving pressure gradient. Equation (\ref{eq:1d-ode-poreFlow})
is exact for isotropic structures and a reasonably good approximation also for anisotropic
porous structures (appendix~\ref{app:1d-aniso}).
Two boundary conditions are needed to close the Brinkman model (\ref{eq:1d-ode-poreFlow}). The first condition is set at the interface with the free fluid. There, velocity is set to slip velocity $u_x = u_s$. We obtain $u_s$ directly from DNS simulations to mimic velocity continuity condition. The other condition is set at the centre of the second REV from the bottom. This shift from the physical wall ensures no influence of the bottom wall on the fitted $\mu$. There, velocity is set to $u_x = - K_{xx} p_x / \mu_f$, i.e.~ the interior (Darcy) velocity.
The analytical solution to equation (\ref{eq:1d-ode-poreFlow}) with the prescribed boundary conditions is
\begin{equation}
    u_b = \left(u_s+\frac{K_{xx}}{\mu_f} p_x \right)e^{y/\delta} - \frac{K_{xx}}{\mu_f} p_x.
    \label{eq:b_vel}
\end{equation}
Here, $\delta = \sqrt{\mu K_{xx}}$ is the characteristic length scale of the exponential decay.
%



\begin{figure}
    \centering
    \hspace*{-1.2cm}
    \includegraphics[trim = 0 80 0 80,clip,width=1.2\textwidth]{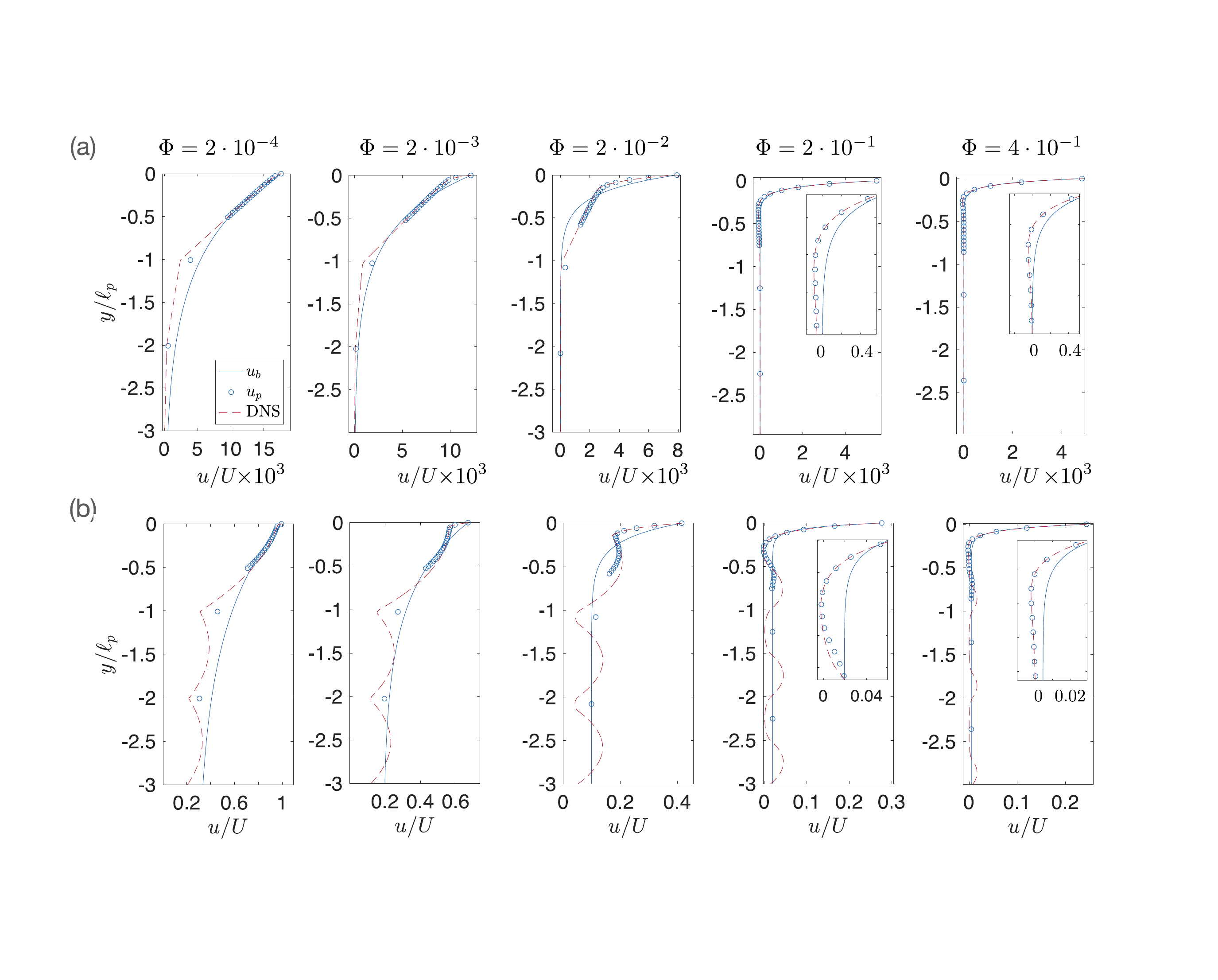}
    \caption{Macroscale velocity profiles ($u_b$), gradually averaged ($u_p$) and plane-averaged velocity profiles from DNS for circular cylinder porous geometries in (a) shear and (b) pressure driven configurations. The insets show negative velocities from DNS and $u_p$.}
    \label{fig:Avg_vel}
\end{figure}


\begin{table}
\begin{center}
\begin{tabular}{p{50mm}p{17mm}p{29mm}}
Geometry & Type & Abbreviation \\ \hline
Circle & Rough & R1 \\
Right-angled triangle & Rough & R2 \\
Equilateral triangle & Rough & R3  \\
Short right-angled triangle & Rough & R4  \\
Square & Smooth & S1  \\
Flipped equilateral triangle & Smooth & S2  \\
Flipped right-angled triangle & Smooth & S3 
\end{tabular}
\end{center}
\caption{Summary of porous geometries evaluated in this work. Abbreviation and type are reported for each geometry. The abbreviation indicates if the geometry is rough or smooth and the sequence number.}
\label{tab:geom-summary}
\end{table}

\subsection{Brinkman viscosity fit}



The Brinkman velocity (\ref{eq:b_vel}) contains one unknown, namely, $\mu$. We use this as a free parameter to obtain the best match between DNS and the Brinkman model. The best match is obtained by minimizing the error in the $2$-norm,
\begin{equation}
    E =  \frac{\|u_{b}-u_{p}\|_2}{\|u_{p}\|_2}. \label{eq:L2error}
\end{equation}
Here,  $u_p$ (the averaged pore-scale solution) and $u_b$ (solution to (\ref{eq:b_vel})) are evaluated at  $y=y_1,\dots,y_N$ with $N=100$.
In figure~\ref{fig:viscosity_example}(c), we present three Brinkman velocity solutions obtained with three different values for $\mu$.
%
We observe that larger $\mu$ reduces the velocity decay rate.
In other words, larger Brinkman viscosity allows more efficient macroscale shear stress transfer towards the interior of the porous media.
Finally, in figure~\ref{fig:viscosity_example}(d) we show the error $E$ as a function of $\mu$, where we observe $E_{\min}\approx 7\%$ at $\mu = 1.81$. The other $\mu$ values (corresponding velocity profiles in figure~\ref{fig:viscosity_example}c) have $E \approx 30\%$.


We apply this minimisation procedure to the porous geometries shown in figure~\ref{fig:PorousModel}(e) for different solid volume fractions.
Velocity distributions for several circular cylinder solid volume fractions are presented in figure~\ref{fig:Avg_vel}. Three velocity profiles are given for each case; the Brinkman solution (blue solid), the plane averaged pore-scale solution (red dashed) and the combined plane and volume-averaged profile (blue circles). In figure \ref{fig:Avg_vel}(a), we show the velocity profile for shear driven configurations.
The plane-averaged DNS profiles exhibit linear regions of velocity decay between the solid structures for small solid volume fractions. For $\Phi \geq 0.2$ recirculation zones appear (figure~\ref{fig:Streamlines}) resulting in a region with negative streamwise velocity (see insets in figure~\ref{fig:Avg_vel}).

In figure \ref{fig:Avg_vel}(b), we show the corresponding pressure-driven velocity profiles. The pore-scale flow field has an oscillatory pattern. The velocity is larger in the void and smaller near the solids, which is a characteristic of Darcy flow. The averaged pore-scale and Brinkman velocity profiles show a uniform velocity in the interior, corresponding to Darcy flow.
Again, for $\Phi \geq 0.2$ recirculation occurs near the interface between the free fluid and porous domain. 

\begin{figure}
    \centering
    \includegraphics[trim = 40 80 40 80,clip,width=\textwidth]{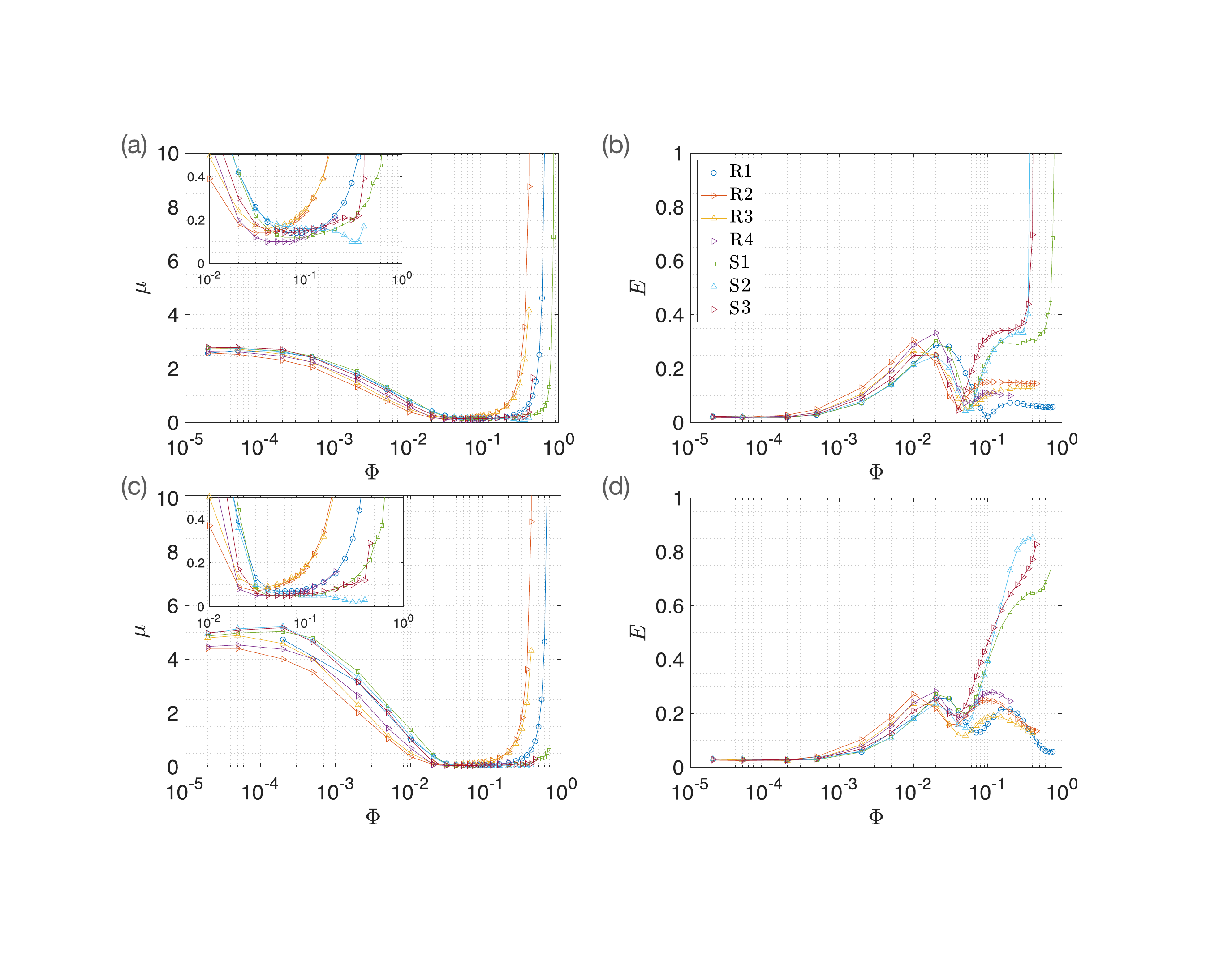}
    \caption{Effective viscosity (a,c) and velocity $L_2$ error norm (b,d) in shear driven (a,b) and pressure-driven (c,d) flow configurations.}
     \label{fig:Mu_summary}
\end{figure}

The best-fit effective viscosity $\mu$ and associated error $E$ are summarized in figure~\ref{fig:Mu_summary}. 
The viscosity ratio $\mu$ as a function of $\Phi$ for shear driven simulations is reported in figure~\ref{fig:Mu_summary}(a).
We observe that $\mu$ obtains values both larger and smaller than unity. For small $\Phi$, all geometries have $\mu\approx 2.4$.
The viscosity ratio $\mu$ begins to be geometry dependent
for $\Phi > 10^{-1}$, where a sharp increase in effective viscosity is observed. The square, flipped equilateral triangle and flipped right triangle (S1-S3) geometries show an increase in effective viscosity for larger $\Phi$ compared to the other geometries. The corresponding $E$ as a function of $\Phi$ is shown in figure~\ref{fig:Mu_summary}(b). The error of the velocity profile for small $\Phi$ is of the order of $10^{-2}$. There is an intermediate maximum in $E$ that occurs around $\Phi \approx 10^{-2}$.
%
For large $\Phi$, two distinct trends in $E$ are observed. An unbounded growth of $E$ with increasing $\Phi$ is observed for the S1-S3 geometries. The remaining geometries (R1-R4) show saturation of $E$. The best-fit viscosity $\mu$ for pressure-driven flows is shown in figure~\ref{fig:Mu_summary}(c). Trends are the same as for shear-driven flow (figure~\ref{fig:Mu_summary}a). However, for small $\Phi$ the viscosity saturates at $\mu\approx 4.5$. The error $E$ for pressure-driven simulations is reported in figure~\ref{fig:Mu_summary}(d). Again, there is an intermediate peak in $E$ for $\Phi \approx 10^{-2}$. For larger $\Phi$, the error $E$ exhibits growth for S1-S3 geometries and reduction for R1-R4. This is qualitatively the same as observed in shear driven simulations (figure~\ref{fig:Mu_summary}b).

\section{Porous surface classification and scaling}\label{sec:Surface}

In this section, we propose a classification of the considered porous geometries.
First, we report the slip length for all structures. Based on the slip length, we divide the geometries into smooth and rough. Then we discuss the scaling of Brinkman viscosity
for the rough porous media.

\subsection{Smooth and rough porous surfaces} \label{sec:categ} 



We classify geometries presented in this study into two categories. To start, we can characterize porous structures visually. Consider the geometries in the right column of figure~\ref{fig:PorousModel}(e), S1-S3.
The top surface of the solid inclusions (that face the free fluid) is flat and aligned with the $x$-axis. 
One may argue that these porous structures are ``smooth''. At the largest $\Phi$, for which solids do not overlap, the structures touch each other. Consequently, these geometries would form a smooth, uninterrupted solid surface. The smooth porous surfaces exhibited unbounded growth of velocity profile error (figure~\ref{fig:Mu_summary}b,d).
Second, we look at the geometries in the left column of figure~\ref{fig:PorousModel}(e), R1-R4. The surfaces at the top do not align with the $x$-axis and we may refer to these materials as ``rough''. In the limit of maximum $\Phi$, these geometries form an impermeable rough surface. The error $E$ for rough structures exhibit a saturation (figure~\ref{fig:Mu_summary}b,d). For each geometry, we report the corresponding category in table~\ref{tab:geom-summary}.

\begin{figure}
    \centering
    \includegraphics[trim = 0 120 0 100,clip,width=\textwidth]{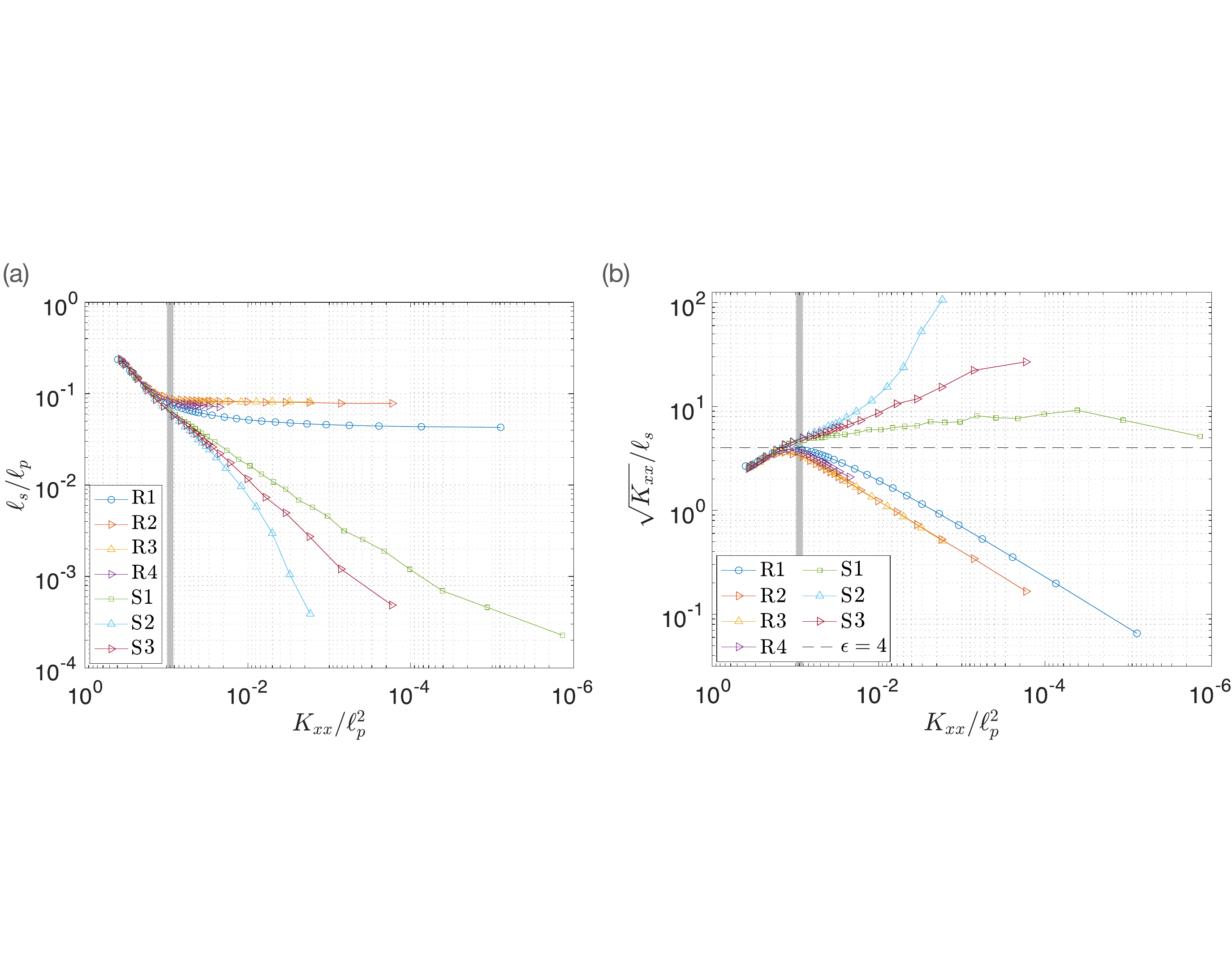}
    \caption{Slip length (a) computed from shear-driven DNS as a function of streamwise permeability. Scale separation $\epsilon = \sqrt{K_{xx}} / \ell_s$ as a function of streamwise permeability (b).}
    \label{fig:slip_kxx_sum}
\end{figure}

A more quantitative distinction between rough and smooth porous surfaces is provided through the Navier-slip length, $\ell_s$, 
%
%
which is defined by the Navier-slip condition, $u_s = \ell_s \frac{d u}{d y}|_{y = 0^+}$.
The slip length can be computed from an interface cell problem~\citep{lacis2016framework}, but here we  extract it from computing the slip velocity $u_s$ and the normal shear $du/dy$ from DNS.
Figure~\ref{fig:slip_kxx_sum}(a) shows $\ell_s$ as a function of permeability $K_{xx}$. 
To have an increasing $\Phi$ on the $x$-axis,
we have chosen to represent decreasing $K_{xx}$ on the $x$-axis. We observe that the smooth surfaces have a continuously decaying slip length (approaching a smooth no-slip surface). On other hand, rough surfaces exhibit a limiting slip value as permeability tends to zero. For large permeabilities (small solid volume fractions, $\Phi < 10^{-2}$) there is no clear distinction between the two types. The vertical grey area in figure~\ref{fig:slip_kxx_sum} show transition in $\Phi$. All data to the left have $\Phi < 10^{-2}$. Contrary, all data to the right have $\Phi > 10^{-2}$. The $K_{xx}$ value at $\Phi = 10^{-2}$ differ between various porous geometries. Therefore finite thickness region is required to define this boundary, corresponding to the maximum and minimum $K_{xx}$ value at $\Phi = 10^{-2}$.

To define a criterion for distinguishing between the two categories,
we propose using a particular scale separation. We define it as a ratio between the permeability associated length scale and
the slip length $\epsilon = 
\sqrt{K_{xx}}/\ell_s$. The scale separation $\epsilon$ as a function of permeability $K_{xx}$
is shown in figure~\ref{fig:slip_kxx_sum}(b).
Here, $\epsilon \approx 4$ seems to be a suitable boundary
between the two types of surfaces.
The porous
surface is considered rough if $\epsilon < 4$ and smooth otherwise. 
Note that $\epsilon < 4$ can be rewritten in terms of effective viscosity estimate \eqref{eq:scale}. By using $C = 1$, we have $\mu = \ell_s^2 K_{xx}^{-1} = \epsilon^{-2}$.
Then $\epsilon < 4$ is equivalent to $\mu^{-1/2} < 4$ or $\mu > 1/16$.
%


\begin{figure}
    \centering
    \includegraphics[trim = 0 140 0 100,clip,width=\textwidth]{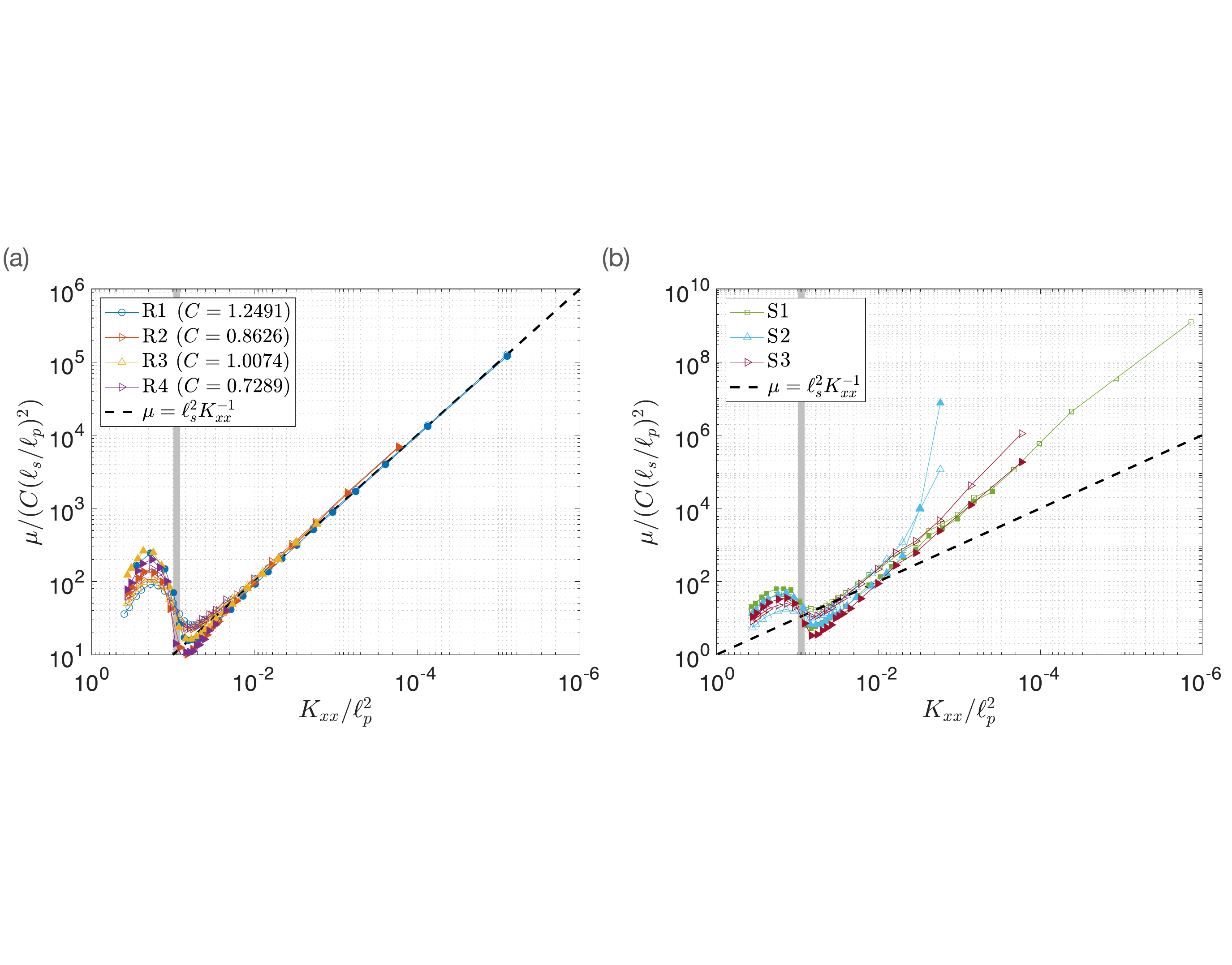}
    \caption{Effective viscosity of rough surfaces (a) and smooth surfaces (b). Open symbols correspond to shear driven flows and closed symbols -- to pressure driven flows.}
     \label{fig:Rough_smooth_scale_mu}
\end{figure}

\subsection{Scaling of the Brinkman viscosity}


The scaling estimate $\mu = C \ell_s^2 K_{xx}^{-1}$ for effective viscosity (\ref{eq:intro-est}) was obtained
in \S\ref{sec:intro} by balancing Brinkman and Darcy terms.
In figure~\ref{fig:Rough_smooth_scale_mu}, 
we present the viscosity ratio $\mu$ normalised by $C (\ell_s / \ell_p)^2$ for all geometries.
%
For the rough surfaces, shown in figure~\ref{fig:Rough_smooth_scale_mu}(a), 
%
 we observe that scaled viscosity curves show good collapse when slip length is selected as the relevant length scale.
The data collapse deteriorates as $\Phi \approx 10^{-2}$ is approached. The proportionality constants required for collapse are $C \sim \mathcal{O}(1)$ for all considered geometries. The exact coefficient values are provided in the legend of figure~\ref{fig:Rough_smooth_scale_mu}(a). On the other hand, the porous surfaces classified as smooth do not collapse with the same scaling (figure~\ref{fig:Rough_smooth_scale_mu}b). The relevant length scale that would allow for an estimate of effective viscosity for smooth surfaces has not yet been identified and is left for future work. 



\section{Discussion}\label{sec:Disc}

In this section, 
we discuss the accuracy one can expect by using the Brinkman model following the guidelines presented in this work. We also put our work in the context of the theoretical limit of $\mu = 1$. And finally, we discuss  issues related to the universality of the Brinkman model.

\subsection{Guidelines for using Brinkman equation}

\begin{figure}
    \centering
    \includegraphics[trim = 0 120 0 120,clip,width=\textwidth]{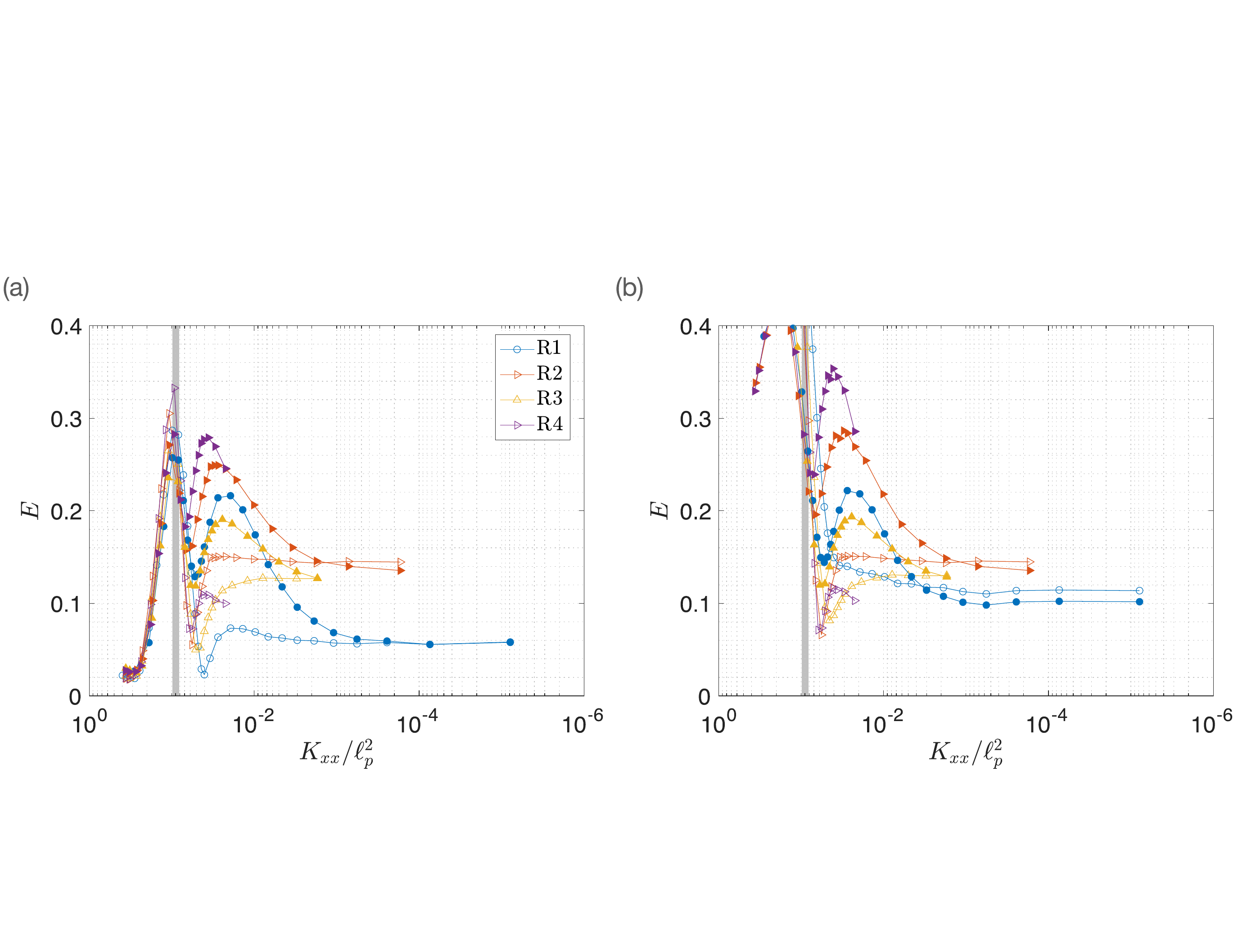}
    \caption{Error of the velocity profiles obtained with (a) best fit effective viscosity and (b) the effective viscosity scaling in equation (\ref{eq:scale}) with $C=1$. Open symbols are shear driven and closed symbols are pressure-driven cases. In (b), the range of $E$ is set equal to (a) for ease of comparison.}
    \label{fig:E2_cunity}
\end{figure}

We propose to estimate the effective viscosity in the Brinkman model from the presented scaling estimate $\mu = C \ell_s^2 K_{xx}^{-1}$ (\ref{eq:intro-est}).
Input parameters are slip length and permeability. These quantities are readily available from simple unit cell computations~\citep{lacis2016framework} or experiments.
%
If possible, the proportionality constant $C$ could be obtained by fitting the Brinkman model to experiments for one value of $\Phi$.
%
If the fit is prohibitive, one may assume $C = 1$.
To justify the use of $C = 1$, we compare the errors associated with the best fit viscosity and the viscosity obtained through the scaling law $\mu = \ell_s^2 K_{xx}^{-1}$ in figure~\ref{fig:E2_cunity}. We observe that the choice $C = 1$ does not lead to a significant increase in the magnitude of the error. When it comes to practically relevant porous media ($\Phi>0.01$),
$E\approx 10\%-15\%$ for pressure-driven configurations while for shear-driven configurations maximum errors are around $30\%-35\%$. 

\begin{figure}
   \centering
   \includegraphics[trim = 0 0 0 0,clip,width=0.55\textwidth]{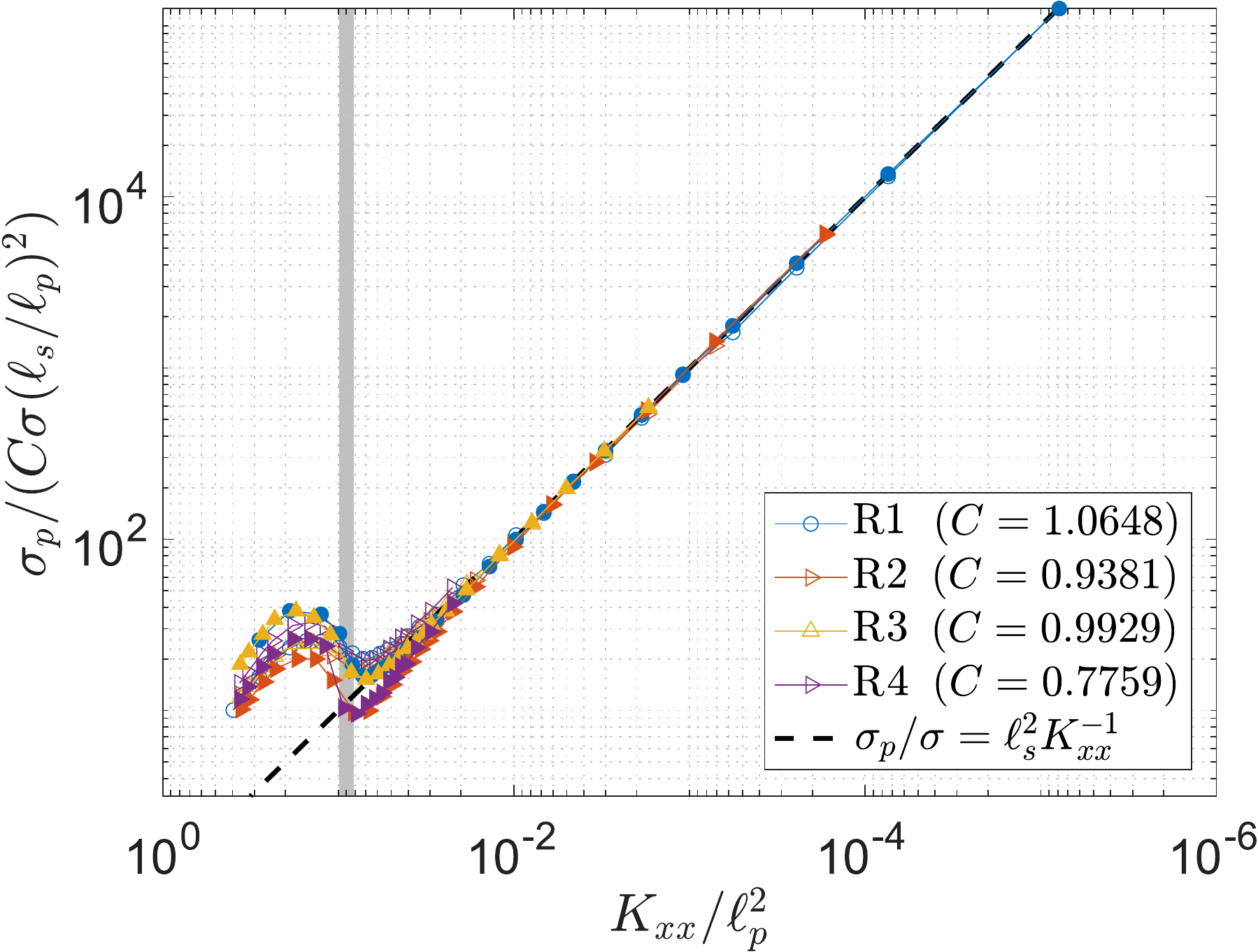}
   \caption{The ratio of interface stress in the porous domain ($\sigma_p$) to free-fluid ($\sigma$). The proportionality coefficients for scaling are provided in the legend.}
    \label{fig:stress_ratio}
\end{figure}

We have assumed velocity continuity between the porous media and the free fluid. To fully define the coupled problem, a stress boundary condition is also required. Although this boundary condition is not investigated in detail in the current work, in what follows, we discuss our results in the context of stress condition.
A stress jump condition can be written as $\ten{\sigma}_p \cdot \vec{n} = 
\ten{R} \cdot \left(\ten{\sigma} \cdot \vec{n}\right)$.
Here, $\ten{\sigma}_p$ is the pore fluid stress tensor,
$\ten{\sigma}$ is the free fluid stress tensor and $\ten{R}$ is a stress ratio 
tensor defining
the free fluid stress transfer to pore fluid~\citep{lacis2019noteTRelast}.
Stress ratio tensor of unity $\ten{R} = \ten{I}$ corresponds to
stress continuity. For the one-dimensional channel flow,
we have only the tangential stress component and the stress jump condition simplifies to
\begin{equation}
    \mu_b \frac{d u_b}{d y} = R_{xx}\,\mu_f \frac{d u}{d y}. \label{eq:tan-str-cond}
\end{equation}
To predict the stress ratio $R_{xx}$, we estimate the normal shear in
free fluid and pore fluid. We focus on the shear-driven configuration. The free-fluid shear can be
estimated using slip length and slip velocity as
\begin{equation}
\frac{d u}{d y} \sim \frac{u_s}{\ell_s}. \label{eq:dudy-ff-est}
\end{equation}
The pore velocity shear can be estimated from the derivative of
the analytical solution (\ref{eq:b_vel}),
\begin{equation}
\frac{d u_b}{d y} \sim \frac{u_s}{\delta}. \label{eq:dudy-pf-est}
\end{equation}
Finally, we estimate the decay length $\delta$ in the Brinkman solution by using the proposed viscosity scaling
\begin{equation}
\delta = \sqrt{\mu\,K_{xx}} \sim \sqrt{ \frac{\ell_s^2}{K_{xx}}\,K_{xx}} = \ell_s. \label{eq:delta-ell}
\end{equation}
Here, it is important to recognise that $\delta$ depends only on the slip length $\ell_s$. This is reasonable because the proposed scaling applies only to rough porous surfaces. Near the interface with the free fluid, rough surfaces can be characterised by slip length $\ell_s$ alone.
This is a similar observation as done by \citet{zaripov2019}. They argued that $\lambda^2 = k\,\mu$ is roughly constant. In our notation, this corresponds to $\delta^2 = \ell_s^2$ being constant. 
Similar exponential decay of velocity profile has been used previously to model the flow above rough surfaces~\citep{liou2009rough,yang2016exponential}.
We further note the following important consequence of (\ref{eq:delta-ell}): For very dense materials, the permeability tends to zero, $K_{xx} \rightarrow 0$. In order to ensure constant $\delta = \ell_s$ for such materials, $\mu$ has to be very large. In the limit of rough impermeable surface ($K_{xx} \rightarrow 0$), $\mu \rightarrow \infty$ is required to obtain $\delta = \ell_s$. Consequently, large $\mu$ values can be expected for rough porous surfaces. This explains the exponential increase of $\mu$ observed in this work (figures~\ref{fig:lit_mu}b;\ref{fig:Mu_summary}a,c;\ref{fig:Rough_smooth_scale_mu}a).

By inserting (\ref{eq:delta-ell}) in (\ref{eq:dudy-pf-est}) and then inserting (\ref{eq:dudy-ff-est},\ref{eq:dudy-pf-est}) into (\ref{eq:tan-str-cond}) we obtain
\begin{equation}
    R_{xx} = \frac{\sigma_p}{\sigma} \sim \mu \sim \frac{\ell_s^2}{K_{xx}}.
    \label{eq:Rxx}
\end{equation}
Here, we have used $\sigma_p = \mu_b\, d u / d y$ and $\sigma = \mu_f\, d u / d y$.
From equations \eqref{eq:Rxx} and \eqref{eq:tan-str-cond}, we observe that
a continuous velocity derivative is predicted.
We evaluate $\sigma_p$ from the Brinkman model with best fit viscosity and compute $\sigma$ from DNS. Finally, we plot the quotient $\sigma_p / ( C\,\sigma\,\ell_s^2)$
as a function of $K_{xx}$ in figure~\ref{fig:stress_ratio}. Here, $C$ again is a proportionality constant of order one. We can
confirm the scaling and the continuity of the velocity derivative.

\subsection{Modelling error}


In this section, we discuss the accuracy one can expect when using the Brinkman model.
There are two main sources for error (figure~\ref{fig:Mu_summary}b,d). 
The first is a step-wise change in the slope of linear velocity segments (figure~\ref{fig:Avg_vel}a, $\Phi  = 0.02$).
The exponential form of the Brinkman solution (\ref{eq:b_vel}) cannot accommodate these step changes using a single decay rate. This results in a Brinkman velocity profile that is smeared out and an over- or under- prediction of the pore-scale solution.
The second source of error is the recirculation that may occur near the interface.
Consequently, there is a negative streamwise velocity (figure~\ref{fig:Avg_vel}a, $\Phi = 0.4$). The exponential form of the Brinkman solution is not capable of capturing negative velocity. This introduces an inevitable error in the modelled velocity profile.
%


\begin{figure}
    \centering
    \includegraphics[trim = 0 130 0 140,clip,width=\textwidth]{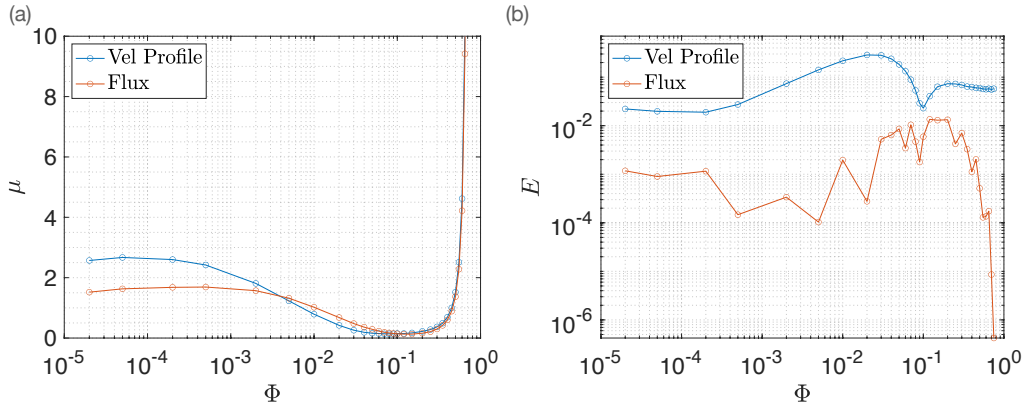}
    \caption{Effective viscosity (a) and modelling error (b) obtained through minimising velocity profile error (\ref{eq:L2error}) and flux error (\ref{eq:flux_error}).} 
    \label{fig:flux_results}
\end{figure}

The error definition used in this work (\ref{eq:L2error})
is based on the shape of the velocity distribution. Alternative approaches for selecting the effective viscosity are possible. For example, one can minimize the error in flux
\begin{equation}
    E = \frac{\int_d^0{u_b \rm{d}y}-\int_d^0{u_p \rm{d}y}}{\int_d^0{u_p \rm{d}y}}.
    \label{eq:flux_error}
\end{equation}
This approach has been used by \cite{zaripov2019}.
The choice of mass flux error leads to quantitatively different $\mu$ as cancellations in $E$ occur. Consequently, the error in mass flux is expected to be smaller than the error in velocity distribution.
To illustrate this, in figure~\ref{fig:flux_results} we present $\mu$ for 
circular cylinders obtained by using the two error definitions.
We observe  similar trends in effective viscosity (figure~\ref{fig:flux_results}a) . The minimisation of the mass flux  allows for a much smaller modelling error (figure~\ref{fig:flux_results}b) compared to minimising \eqref{eq:L2error}. The error is less than $2\%$ over the 
considered $\Phi$ range.


Finally, we discuss the difference between the shear driven and pressure-driven configurations. We observe (figure~\ref{fig:Mu_summary}a,c) insignificant differences in $\mu$ for large $\Phi$. In this regime, $K_{xx}$ is very small. This in turn renders the pressure term in equation (\ref{eq:b_vel}) negligible. In other words, the permeability is so small that the effect of porosity is not felt.
However, for small $\Phi$ values the spread of viscosity data
is more visible (figure~\ref{fig:Rough_smooth_scale_mu}a). There is a
difference between best viscosity for shear and pressure-driven flows. 


\begin{figure}
    \centering
    \includegraphics[trim = 0 180 0 140,clip,width=\textwidth]{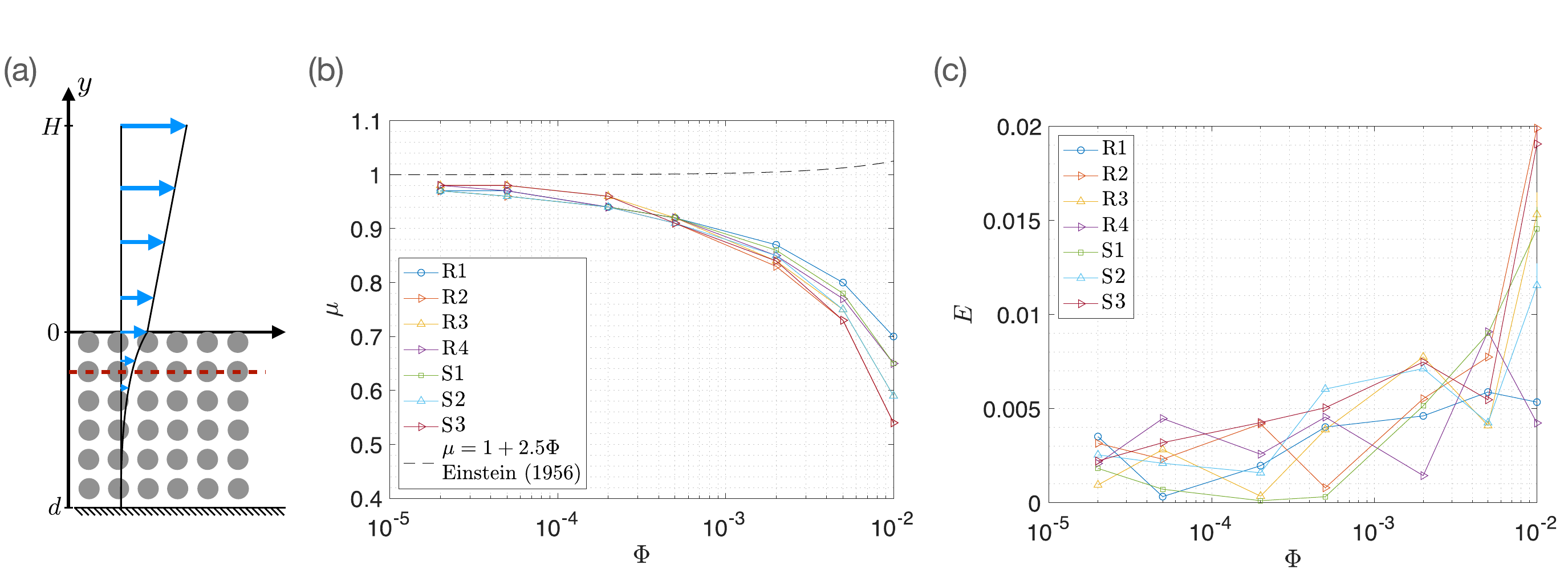}
    \caption{Effective viscosity (b) and modelling error (c) for the interior of porous material. Red dashed line shows the boundary of the interior domain (a).} 
    \label{fig:2d_results}
\end{figure}

\subsection{Brinkman viscosity of unity}


In this section, we discuss the value of effective viscosity $\mu$
as $\Phi$ approaches zero.
Existing theoretical studies~\citep{einstein1956investigations,freed1978,Ochoa-Tapia_1995,Starov2001} all predict $\mu = 1$ in this limit.
%
%
The effective viscosity in figure~\ref{fig:Mu_summary}a,c show a clear deviation from $\mu = 1$. We observe that the asymptotic limits are $\mu = 2.4$ for the shear driven and $\mu = 4.5$ for the pressure-driven configurations.
To understand the cause of this difference, we investigate the flow only in the interior of the porous media. With a dashed red line in figure~\ref{fig:2d_results}(a), we show the boundary below which we expect only interior flow.
We truncate the Brinkman solution (\ref{eq:b_vel}) to exclude the region closest to the interface.
We impose the averaged velocity measured from the shear driven DNS at the red dashed
line (figure~\ref{fig:2d_results}a) as the boundary condition for the truncated Brinkman
solution. Then, we find $\mu$ for the interior by carrying out the same fitting procedure as described
in \S\ref{sec:BrinkVisc}. 
The obtained $\mu$ as a function of $\Phi$ is shown in figure~\ref{fig:2d_results}(b). Note that we obtain $\mu$ only for $\Phi < 0.01$. For $\Phi > 0.01$ the velocity below the red dashed line is vanishingly small and obtaining $\mu$ is not possible.
%
Also, note that the obtained variation in $\mu$ exhibits significant
differences from the existing theoretical predictions~\citep{einstein1956investigations,Starov2001,Ochoa-Tapia_1995,freed1978}.
All the literature predictions are indistinguishable from each other in
the solid volume fraction range $10^{-5} < \Phi < 10^{-2}$, therefore we show only one curve
in figure~\ref{fig:2d_results}(b).
%
Despite the difference from theory, the Brinkman equation proves to be an excellent model for the interior domain. In our $\mu$ fit for the interior, we observe very small modelling errors (figure~\ref{fig:2d_results}c). 

\subsection{Irregular geometry}

\begin{figure}
    \centering
    \includegraphics[trim = 0 150 0 120,clip,width=\textwidth]{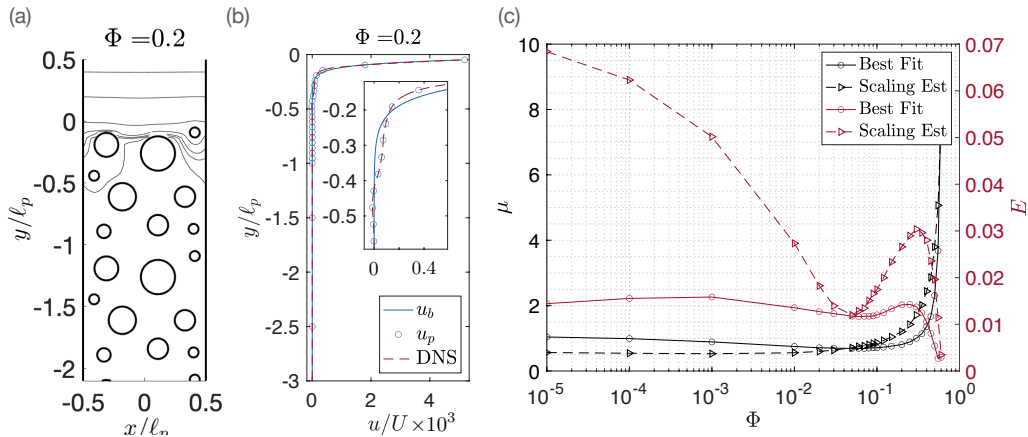}
    \caption{Streamlines (a) and (b) velocity profiles of flow around irregular circular cylinders with $\Phi = 0.2$. Effective viscosity and $L_2$ error norm obtained through best fit and proposed scaling estimate (c).}
    \label{fig:irr_results}
\end{figure}

The recirculation and the linearity of the velocity profile are suppressed in porous geometries that have an irregular structure. To demonstrate this, we consider a porous material constructed from circular cylinders with various radii (figure~\ref{fig:irr_results}).
In figure~\ref{fig:irr_results}(a,b), we present streamlines and the velocity profile for $\Phi = 0.2$. We do not observe any pronounced recirculation. Consequently, there is no negative region in the velocity distribution. The velocity profile for this structure, however, exhibits a much faster velocity decay compared to regular structures. The reason is an increased number of solids within one REV. Despite the rapid decay, the DNS velocity profile is smooth. This profile can be easily captured by the Brinkman model (figure~\ref{fig:irr_results}b). In figure~\ref{fig:irr_results}(c), we show the viscosity ratio $\mu$ and the velocity profile error $E$ over the considered solid volume fractions. The error is significantly smaller than for regular structures. The maximum $E$ for the irregular configuration is on the order of the minimum $E$ for the regular ones. This indicates that the Brinkman model can be more accurate for irregular porous structures. The effective viscosity exhibits similar trends as for the regular cases. There is an asymptotic approach to a constant value in the small $\Phi$ limit. For large $\Phi$, the effective viscosity rapidly increases.

It is interesting to understand if the estimate $\mu = \ell_s^2 K_{xx}^{-1}$ is applicable also for the selected irregular geometry.
The viscosity $\mu$ from the scaling estimate is shown in figure~\ref{fig:irr_results}(c) with black dashed lines. The corresponding error $E$ is shown in figure~\ref{fig:irr_results}(c) with red dashed lines. We observe that the error is larger compared to the error associated with best-fit viscosity. However, it is always below $4\%$ for practically relevant solid volume fractions. This indicates that the scaling $\mu = \ell_s^2 K_{xx}^{-1}$ is likely to be valid also for rough irregular porous structures.
However, additional work is required to draw a more general conclusion.

\begin{figure}
    \centering
    \hspace*{-1.7cm}
    \includegraphics[trim = 20 130 20 140,clip,width=1.18\textwidth]{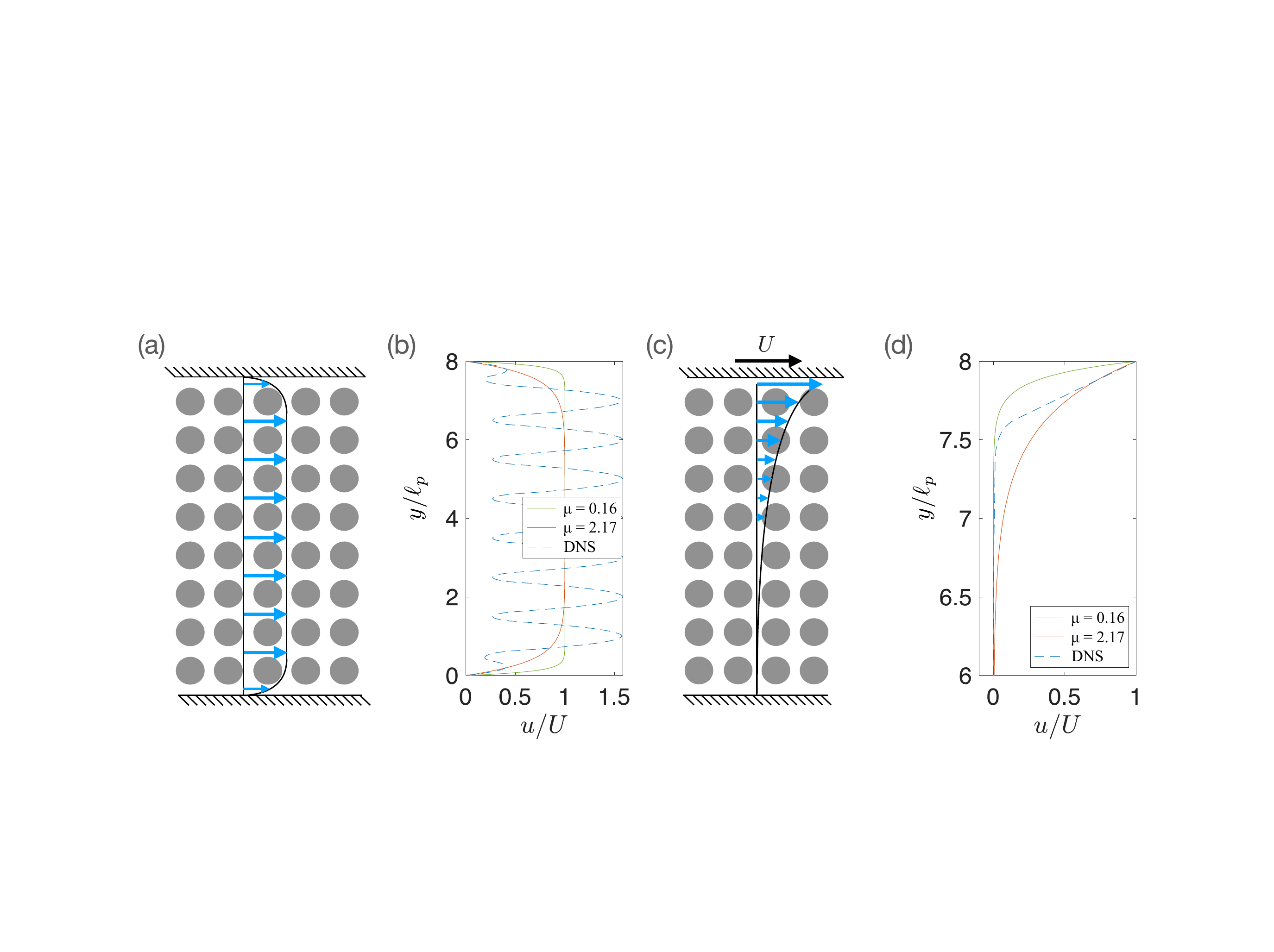}
    \caption{Pressure driven porous channel flow (a) with corresponding pore-scale and Brinkman
    solutions (b) with viscosities $\mu = 2.17$ \citep{zaripov2019} and $\mu = 0.16$ (this work). In (c,d) the flow is driven through shear
    instead of the pressure gradient. Circular cylinders with solid volume fraction $\Phi = 0.05$.} 
    \label{fig:2d_results_int}
\end{figure}

\subsection{Application of the Brinkman model to different flow configurations}

In this section, we apply the Brinkman model to a channel filled with a porous material. First, we consider a pressure-driven flow (figure~\ref{fig:2d_results_int}a). As the porous material, we use a total of $8$ cylinders in the vertical direction at $\Phi = 0.05$. This also results in a one-dimensional flow. The same configuration is investigated by \citet{zaripov2019}. In figure~\ref{fig:2d_results_int}(b) we show the line averaged velocity distribution obtained from DNS and predictions from the Brinkman model. For $\mu$, we have used value fitted by \citet{zaripov2019}, $\mu = 2.17$, and the best-fit value from the current work, $\mu = 0.16$. We compute the error in mass flux and obtain $0.06\%$ for $\mu = 2.17$ and $7.56\%$ for $\mu = 0.16$. This clearly illustrates that different Brinkman viscosity is needed to obtain good mass flux prediction in confined pressure-driven flow through porous media.

To further highlight the dependence of $\mu$ on the system configuration, we change the driving mechanism.
We set the pressure gradient to zero and move the top wall, as shown in figure~\ref{fig:2d_results_int}(c).
The corresponding DNS and Brinkman velocity
profiles with $\mu = 2.17$ and $\mu = 0.16$ are shown in figure~\ref{fig:2d_results_int}(d).
Computing error in velocity flux yields $62.5\%$ for $\mu = 2.17$ and $55.8\%$ 
for $\mu = 0.16$. After the change in the driving mechanism of the flow, we observe
that neither viscosity value provides an acceptable representation of the velocity flux.
Although not demonstrated here, the same sensitivity exists to small changes
in distance between the last pore structure and the confining wall both in pressure
and in shear-driven porous channel flows.




These results serve to demonstrate that the effective viscosity is highly sensitive to the wall location and driving mechanism. This most likely is the culprit that hinders the efforts to obtain universal Brinkman viscosity for a specific porous geometry.
One possible reason for this sensitivity could be the value of the permeability near the interface. It has previously been shown that the permeability at the interface with free fluid differs from that of interior~\citep{lacis2016framework}. A similar effect could also occur near a solid wall. Another possibility is a limitation of the Brinkman model as an ad-hoc description of macroscale flow. 
Nevertheless,  for a specific geometry and flow configuration, it seems that there exists a $\mu$ that provides relatively low modelling errors.
In this work, we provide $\mu$ for free flow over porous material. For pressure-driven porous channel flow, $\mu$ can be found in the work by \citet{zaripov2019}. For flow configuration different than these, a separate investigation would be required. 


\section{Conclusions}\label{sec:Conc}

In this work, we have proposed to estimate the Brinkman viscosity by $\mu = \ell_s^2 / K_{xx}$. We have shown that the Brinkman model with the proposed $\mu$ is suitable to model flows over and through rough porous media. We have provided a quantitative criterion $\sqrt{K_{xx}}/\ell_s < 4$ (equivalent to $\mu > 1/16$) to distinguish between rough and smooth porous media. The interface between the free fluid and porous media was chosen at the crest plane (the tip of the last solid structures).
By considering several different porous geometries, we have demonstrated that the expected error of the Brinkman model for rough porous surfaces remains bounded below approximately $30\%$, while for smooth porous surfaces the modelling error grows without bound.

We have found that the main cause of error for intermediate $\Phi \sim 0.02$ is the piece-wise linear behaviour of the DNS velocity profile. On other hand, for larger $\Phi \sim 0.1$ the main cause of the error is recirculation and negative stream-wise velocity. Both of these effects can not be captured by a simple exponential decay predicted by the Brinkman model. These effects are relevant to the large class of regular porous materials, such as those obtained through 3D printing. We have also demonstrated on one irregular porous geometry that the velocity profile is smoothed and the recirculation is suppressed. Consequently, the modelling error of the Brinkman equation is much smaller. The viscosity scaling $\mu = \ell_s^2 / K_{xx}$ allowed to obtain effective velocity profiles with error below $4\%$ for $\Phi > 0.01$.

We have further discussed the consequences of our results for the stress condition at the interface between free fluid and porous material. We have demonstrated that the continuity of $du / dy$ seems the most appropriate boundary condition. In addition, we have noticed that for rough surfaces the exponential decay length is equal to slip length $\delta = \ell_s$. This leads to very large $\mu$ for rough surfaces with very small permeability. 
We have also discussed the approach of the Brinkman viscosity to fluid viscosity $\mu = 1$ as solid volume fraction approach zero. We have obtained this for interior flow only. Finally, we have compared the Brinkman model and DNS with the $\mu$ values obtained in this work and proposed by \citet{zaripov2019} for porous channel flow. Neither $\mu$ value has been successful in capturing flow distribution for the configuration with a moving wall.

Since the accuracy and applicability of the Brinkman model is configuration-specific, further work is required on a range of canonical problems. For the particular configuration of regular rough porous media in contact with free fluid, we have provided a consistent and robust way to determine Brinkman viscosity value.
We believe that this work will serve as valuable insight for practitioners that want to continue using the Brinkman model for solving  applied flow problems
in biology, geology and engineering.



\appendix

\section{One-dimensional flow for anisotropic porous structures} \label{app:1d-aniso}

In this appendix, we demonstrate the full one-dimensional equations
for anisotropic structures. We motivate why the same equation as
for isotropic porous structures is a sufficiently good approximation.
We expand the Brinkman equation (\ref{eq:Brinkman}) for one-dimensional
flow over and through an anisotropic porous material. We get 
\begin{equation}
u_x = - \frac{K_{xx}}{\mu_f} \frac{\partial p}{\partial x} - \frac{K_{xy}}{\mu_f} \frac{\partial p}{\partial y} + \frac{\mu_b}{\mu_f} K_{xx} \frac{d^2 u_x}{dy^2}. \label{app:eq:aniso-stream}
\end{equation}
This must be coupled with the condition for vertical pressure gradient
\begin{equation}
0 = - \frac{K_{yx}}{\mu_f} \frac{\partial p}{\partial x} - \frac{K_{yy}}{\mu_f} \frac{\partial p}{\partial y} + \frac{\mu_b}{\mu_f} K_{yx} \frac{d^2 u_x}{dy^2}.
\end{equation}
The horizontal pressure gradient $\partial p / \partial x$ can
be understood as a constant externally imposed driving pressure gradient.
The vertical pressure gradient $\partial p / \partial y$ can be viewed as
a constant constraint ensuring no wall-normal velocity. From the constraint
equation, we express the vertical pressure gradient as
\begin{equation}
\frac{\partial p}{\partial y} = - \frac{K_{yx}}{K_{yy}} \frac{\partial p}{\partial x}  + \mu_b \frac{K_{yx}}{K_{yy}} \frac{d^2 u_x}{dy^2}. \label{app:eq-vert-p-grad}
\end{equation}
Furthermore, we use the symmetry of the permeability
tensor $K_{xy} = K_{yx}$. Then we insert the vertical pressure gradient (\ref{app:eq-vert-p-grad}) in
the equation for streamwise velocity (\ref{app:eq:aniso-stream}). This gives
\begin{equation}
u_x = - \frac{1}{\mu_f} \left( K_{xx} - \frac{K^2_{xy}}{K_{yy}} \right) \frac{\partial p}{\partial x} + \frac{\mu_b}{\mu_f} \left( K_{xx} - \frac{K^2_{xy}}{K_{yy}} \right) \frac{d^2 u_x}{dy^2}.
\end{equation}
We observe that the effect from the anisotropy of the porous material exhibits itself
for the 1D flow as a correction to streamwise permeability. 
However, the off-diagonal permeability term typically is much smaller than
the diagonal terms, $K_{xy} \ll K_{xx}$ and $K_{xy} \ll K_{yy}$. We observe
that the correction of the streamwise velocity due to the anisotropy
of the porous material is quadratic in $K_{xy}$. Therefore it is safe to
neglect the correction and
use the equation (\ref{eq:1d-ode-poreFlow}) obtained for isotropic
structures even for
anisotropic structures.

\begin{acknowledgments}
A.R. S.B. and U.L. were funded through the Knut and Alice Wallenberg Foundation (KAW 2016.0255), the Swedish Foundation for Strategic Research (SFF, FFL15:0001) and Swedish Research Council (VR-2014-5680, INTERFACE centre). We acknowledge Samuel Sirot for carrying out the preliminary work in this project during his internship time at Department of Engineering Mechanics, KTH.
\end{acknowledgments}

\bibliographystyle{jfm}
\bibliography{Brinkman.bib}

\end{document}